\documentclass{webofc}
\usepackage{here}
\usepackage[varg]{txfonts}   
\begin{document}
\title{Coupled dynamics of heavy and light flavor flow harmonics from EPOSHQ}
%
%

\author{\firstname{P.B.} \lastname{Gossiaux}\inst{1}\fnsep\thanks{\email{gossiaux@subatech.in2p3.fr}} \and
 \firstname{J.} \lastname{Aichelin}\inst{1} \and
 \firstname{M.} \lastname{Nahrgang}\inst{1}\and
 \firstname{V.} \lastname{Ozvenchuk}\inst{1,2}\and
 \firstname{K.} \lastname{Werner}\inst{1}}

\institute{SUBATECH, UMR 6457, IMT Atlantique, Universit\'e de Nantes, IN2P3/CNRS.\\ 4 rue Alfred Kastler, 44307 Nantes cedex 3, France
\and
 Cracow, INP, ul. Radzikowskiego 152, 31-342 Krak\'ow, Poland}

\abstract{We pursue the study of event by event correlations between low-mass particles and heavy mesons flow harmonics in ultrarelativistic heavy ion collisions and clarify some ambiguities found in one of our previous work.}
\maketitle
\section{Introduction and method}
\label{intro}
The joint study of light and heavy particles produced in ultrarelativistic heavy ion collisions offers new insights on several physical aspects, in particular on the coupling of heavy quarks with the quark gluon plasma formed in those collisions. In \cite{Nahrgang:2015}, we have focused on the comparison between flow harmonics of light and heavy mesons and conjectured that heavy mesons would benefit less from the bulk flow due to their inertia. Such an effect could f.i. manifest itself in event by event (EBE) correlations between those flow harmonics, as studied recently in \cite{Prado:2016szr} and \cite{gossiauxQM}, where anomalous fluctuations where found.     

In this study, we follow the same method as in \cite{gossiauxQM}, using the EPOSHQ model which allows to describe and predict both the light and the heavy sectors in ultrarelativistic pp, pA and AA collisions. Heavy quarks strongly interact with the QGP phase and thus benefit from the collective flows of the bulk. Later on, D mesons are allowed to interact with light hadrons in the hadronic phase, leading to some moderate extra flow. For the sake of simplicity in this seminal work with the EPOSHQ model, D mesons are assumed to be produced exclusively through fragmentation mechanism. The analysis is performed EBE using the event plane method, which does not suffer from the usual caveats as the participant plane is known with a very good accuracy from the initial state conditions. 
In \cite{gossiauxQM}, we have adopted an artificial cranking up of the c quark production by a factor 50 in order to reduce the $v_n$ fluctuations resulting from the limited number of c quarks~\cite{Poskanzer:1998}. This however seems to lead to abnormally large fluctuations and even led us to conjecture the existence of large EBE fluctuations specific to heavy flavor production and propagation in QGP. In this work, we do not resort to such an oversampling means and provide brute results for D mesons, at the cost of an extra computational effort.    

\section{Results and analysis}

In fig.~\ref{fig-1}, we show the distributions of the $\hat{v}_{n=2,3}$ obtained\footnote{According to $\hat{v}_n\{\rm PP\}=\frac{1}{N}\sum_{i=1}^N \cos(n(\phi_i-\hat{\psi}_n^{\rm PP}))$, where $N$ is the number of selected particles in the sample.} for $\pi$, p, c-quarks (before hadronization) and 
D mesons in EPOSHQ events, taking $\Psi_n^{\rm PP}$ as the reference plane for all species. The large variances observed for the flows of nucleons and especially for HF mesons imply that the{ir individual flows extracted EBE are often found with opposite sign as compared to the average.  
\begin{figure}[H]
	\centering
	\includegraphics[width=10cm,clip]{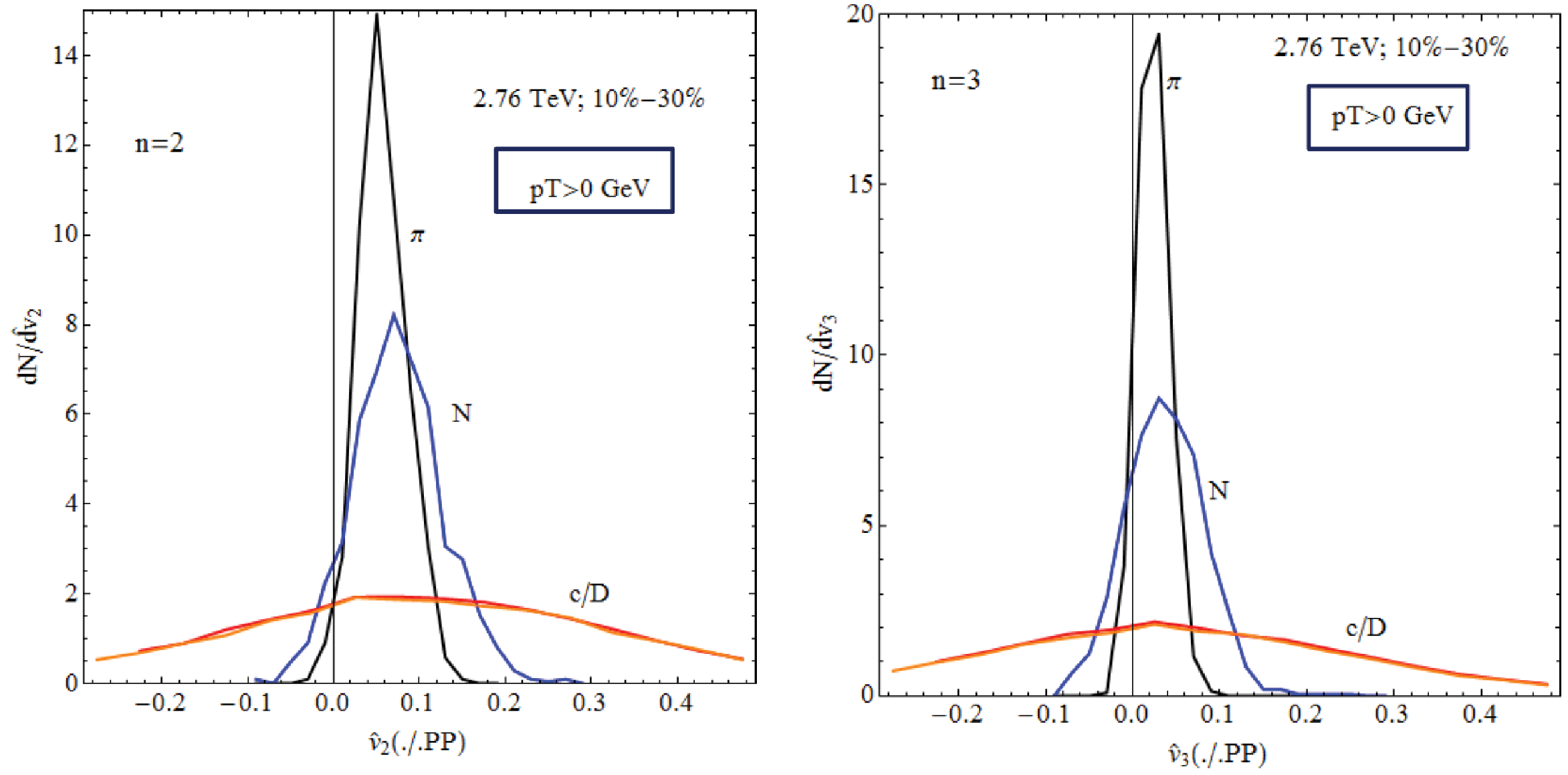}
	\caption{The distributions of elliptic (left) and triangular (right) flows of final pions, nucleons (N), charmed quarks and D mesons produced in Pb-Pb collisions at $\sqrt{s}=2.76~{\rm TeV}$ in the 10\%-30\% centrality class.}
	\label{fig-1}       
\end{figure}
Following~\cite{Poskanzer:1998}, we expect the variance of $\hat{v}_{n=2,3}$
to be 
\begin{equation}
{\rm var}(\hat{v}_{n})\approx \frac{1}{2 N}+{\rm var}_{{\rm EBE},n}
\label{eq:var}
\end{equation}
In table~\ref{tab-1}, we provide the numbers associated to the various cases shown in fig.~\ref{fig-1}. We observe a clear transition from $\pi$ (dominance of EBE fluctuations) $\rightarrow {\rm N} \rightarrow {\rm D}$, for which the fluctuations are quite well explained by the usual $1/2N$ term in eq.~(\ref{eq:var}), in contrast with our previous observation~\cite{gossiauxQM}. We have checked that this holds for other centrality classes, other energies, larger HF coupling to the QGP bulk, finite $p_T$ cut off, etc. 
\begin{table}[H]
	\centering
	\caption{Decomposition of the flow variance for the 10\%-30\% centrality class.}
	\label{tab-1}       
	\begin{tabular}{ccccc}
		\hline
		& ${\rm var}(\hat{v}_{2/3})$ & $\langle \frac{1}{2 N}\rangle$ & residual (${\rm var}(\hat{v}_{2/3})-\langle \frac{1}{2 N}\rangle$) &
		$\%$age of fluct. due to $\langle \frac{1}{2 N}\rangle$
		 \\\hline
		$\pi$ & 8.2E-4/3.3E-4 & 1.3E-4 & 6.9E-4/2E-4 & 16\%/39\% \\
		N & 2.9E-3/2.2E-3 & 1.8E-3 & 1.1E-3/4E-4 & 62\%/82\%\\
		D & 8.5E-2/8E-2 & 7.9E-2 & 5.7E-3/1.6E-3 & 93\%/99\%\\\hline
	\end{tabular}
\end{table}
In fig.~\ref{fig-2}, we show the elliptic and triangular flows of D mesons as a function of the spatial eccentricities, in a similar way as our QM2014 approach. Assuming a linear increase of $\langle v_n\rangle$ with $\epsilon_n$ in the bulk part of the event distribution, one obtains ${\rm var}_{{\rm EBE},n} \approx (\langle v_n\rangle/\epsilon_n)^2{\rm var}(\epsilon_n)$, i.e. ${\rm var}_{{\rm EBE},2} \approx$ 6.8E-4/1.2E-3/1.2E-3 for $\pi$/N/D mesons. These numbers should be compared with the third column of table~\ref{tab-1}, with excellent agreement for $\pi$ and nucleons. For HF, an extra source of fluctuations cannot be excluded, but it is definitively much smaller that what was advocated in \cite{gossiauxQM} and is superseded by the $1/2N_D$ contribution for practical purposes.
\begin{figure}[H]
	\centering
\includegraphics[width=10cm,clip]{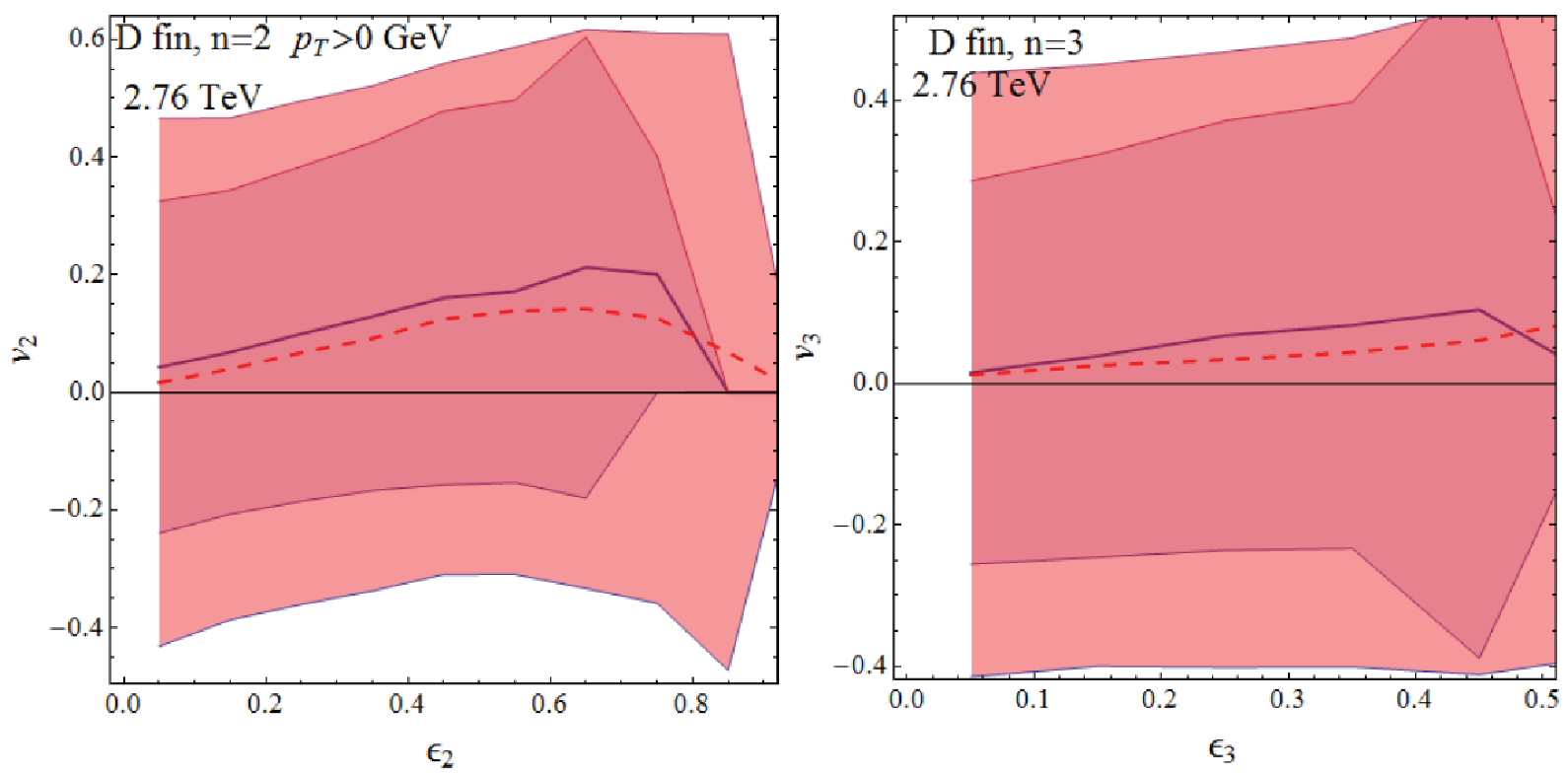}
	\caption{The elliptic (left) and triangular (right) flows of final D mesons produced in Pb-Pb collisions at $\sqrt{s}=2.76~{\rm TeV}$ as a function of spatial eccentricity; the solid lines correspond to average values (plain for 10\%-30\% centrality class and dashed for 30\%-50\% centrality class). The red (30\%-50\%) and darker (10\%-30\%) bands represent the average $\pm 1~\sigma$ interval.}
	\label{fig-2}       
\end{figure}
\begin{figure}[H]
	\centering
		\includegraphics[width=10cm,clip]{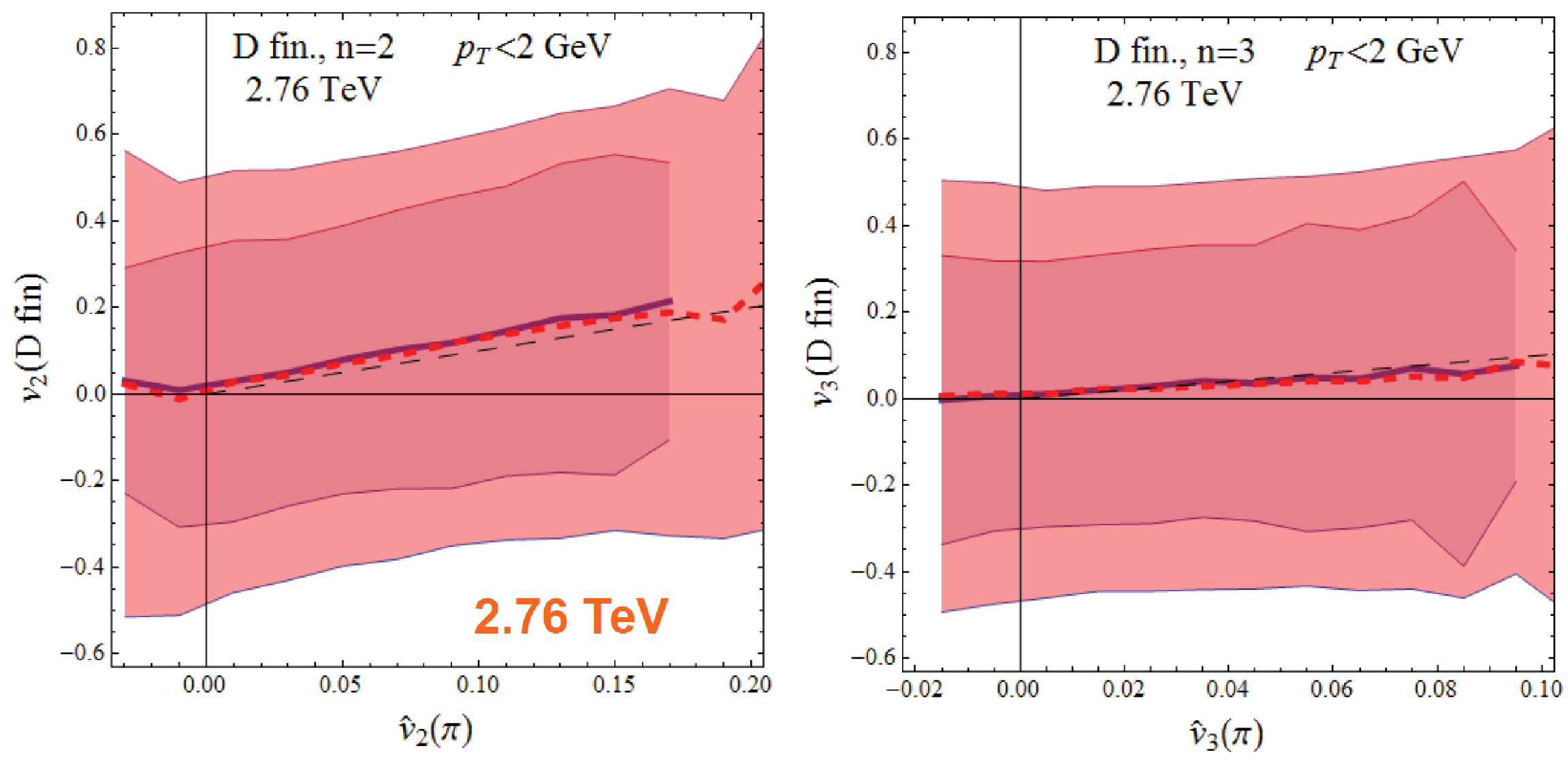}
	\caption{The elliptic (left) and triangular (right) flows of final $p_T<2~{\rm GeV}$ D mesons produced in Pb-Pb collisions at $\sqrt{s}=2.76~{\rm TeV}$ as a function of equivalent quantity for $\pi$; same conventions as for fig.~\ref{fig-2}. The thin dashed line represents the $\langle v_n(D)\rangle =\langle v_n(\pi)\rangle$ case.}
	\label{fig-3}       
\end{figure}
We have conducted a refined analysis to identify the origin of the large fluctuations found previously with the oversampling method. In practice such an oversampling is achieved by performing 50 so-called "HQ events" -- i.e. evolution of the HQ in the bulk -- for a single EPOS3 global event with the {\em same} initial heavy quarks. This is an unavoidable feature as the production of heavy-quarks is deeply rooted in the semi-hard pomeron approach of EPOS3. The 50 replications of the HQ $\hat{v}_n$ are therefore totally correlated at initial time and gradually decorrelate with the time evolution. During our previous analysis, we had taken care to check that a very good level of decorrelation is achieved at the end of the evolution. However, we have realized since then that a small persisting correlation generates, in turn, an extra contribution in ${\rm var}(\hat{v}_n)$ that scales like the number of replications and thus appears as a contribution to ${\rm var}_{{\rm EBE},n}$ in eq.~(\ref{eq:var}). This explains our misinterpretation in \cite{gossiauxQM}. 

Having the fluctuations under control, we focus on the averages $\langle v_n\rangle$ in fig.~\ref{fig-2}. We find similar conclusions at those stemming from our study \cite{Nahrgang:2015} with EPOS2 as a bulk: Heavy quarks benefit less and less from the bulk flows for higher and higher eccentricities, harmonics (and mass). 
In fig.~\ref{fig-3} and~\ref{fig-4}, we present the correlations between the flows of D mesons and the equivalent flows of final pions\footnote{Taken as the best proxy to probe the initial eccentricities of the Pb-Pb collisions.} for low and intermediate $p_T$ particles. In the low $p_T$ case -- where one expects nearly perfect thermalization of heavy quarks --, one finds very good correlations between the light and the heavy sectors. For intermediate $p_T$ particles, the correlation is less pronounced and one observes a saturation for the largest values of $v_n(\pi)$, especially in the 30\%-50\% centrality class.

\begin{figure}[H]
	\centering
	\includegraphics[width=10cm,clip]{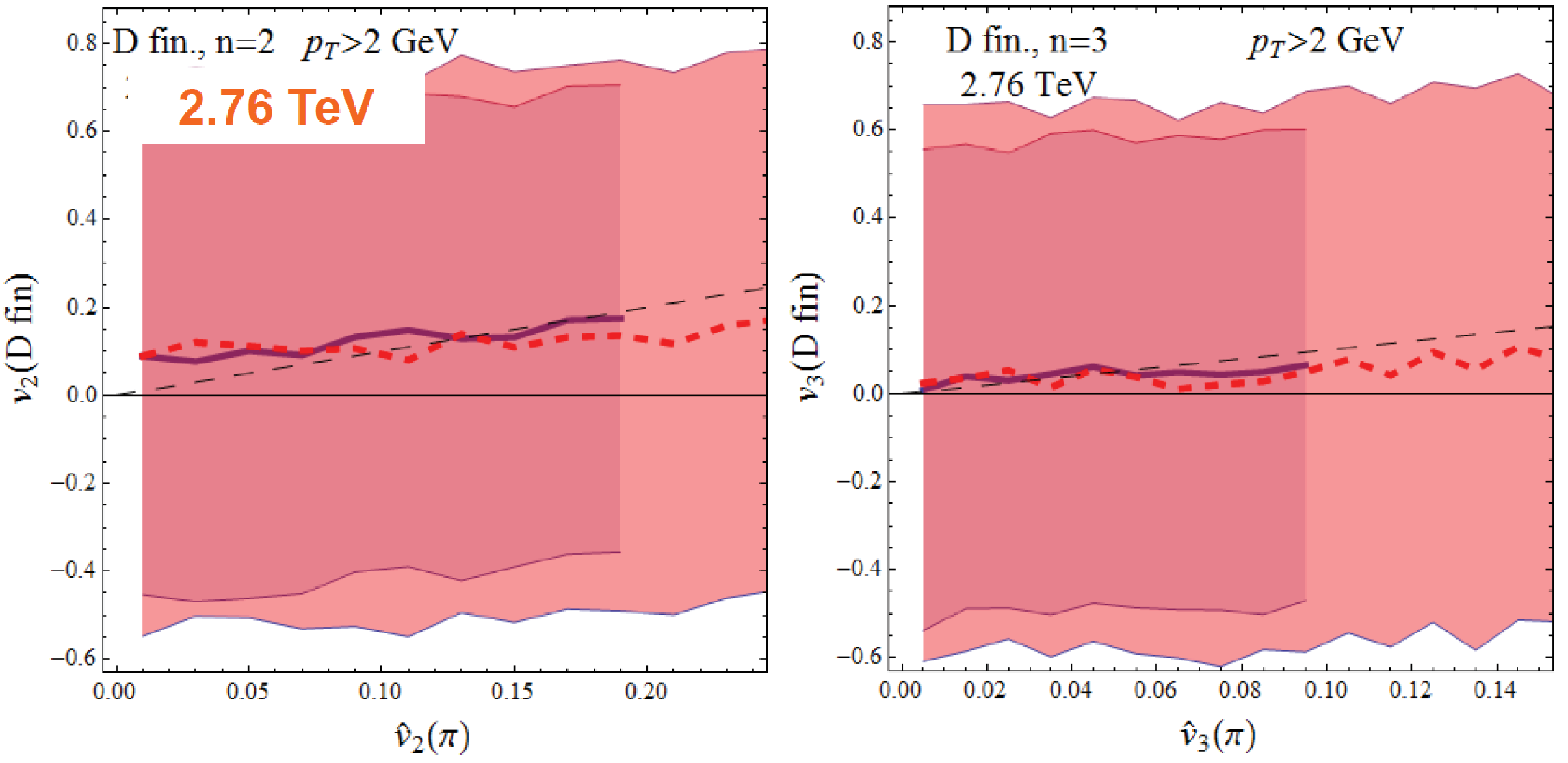}
	\caption{Same as fig.~\ref{fig-3} for $p_T>2~{\rm GeV}$ mesons.}
	\label{fig-4}       
\end{figure}

\section{Conclusions}
We have refined our analysis of heavy-light flow correlations initiated in 
\cite{gossiauxQM} in two main aspects: First, we have understood that the HQ oversampling procedure leads to spurious extra-fluctuations in the EPOSHQ framework and thus clarified the ambiguities raised in \cite{gossiauxQM}. Although some specific EBE fluctuations of the HF harmonic-flows cannot be ruled out, they appear to be subdominant. Second, we have increased our samples and present more robust predictions for the correlations between D mesons and pions harmonic flows. These could be measured experimentally by event engineering techniques, triggering on the $v_n$ of pions either in minimum bias events or in a given centrality class. Going towards larger $\sqrt{s}$ facilitates this study, as the fluctuations are reduced due to the more abundant HQ production.

\section*{Acknowledgments}
We gratefully acknowledge the support from the R\'egion Pays de la Loire (France) as well as enlightening discussions with J.Y. Ollitrault.


\begin{thebibliography}{}
%
%
\bibitem{Nahrgang:2015}
M. Nahrgang et al., Phys. Rev. C{\bf 91},  014904  (2015)
\bibitem{Prado:2016szr}
C.A.G.~Prado et al, arxiv:1611.02965
\bibitem{gossiauxQM}
P.B. Gossiaux et al., Nuc. Phys. A{\bf 967}, 672 (2017)
\bibitem{Poskanzer:1998}
A.~Poskanzer and S.~Voloshin, Phys. Rev. C{\bf 58}, 1671 (1998)

\end{thebibliography}
\end{document}